\documentclass[aps,prc,twocolumn,fleqn,floatfix,nofootinbib]{revtex4}
\usepackage{amsmath,amssymb,amsfonts}
\usepackage{graphicx}
\usepackage{bm}

\newcommand \tie {{i.e.}}

\def\be{\begin{eqnarray}}
\def\ee{\end{eqnarray}}

\def\blambda{{\bar{\lambda}}}

\def\xidmm{ \ln\frac{x_m Q^2_{max}}{\mu^2}}
\def\xisat{\xi_{s}\,}
\def\xidmx{\ln\frac{x^2_m Q^2_{max}}{\mu^2}}
\def\yd{\ln\frac{1}{x}}
\def\lxinc{\ln \left({\ln\frac{Q^2_s}{\Lambda^2}}\Big/{\ln\frac{\mu^2}{\Lambda^2}}\right)}
\def\lxincm{\ln \left({\ln\frac{Q^2_s(x_m)}{\Lambda^2}}\Big/{\ln\frac{\mu^2}{\Lambda^2}}\right)}
\def\lxincR{\ln \left({\ln\frac{Q^2_s+T^2}{\Lambda^2}}\Big/{\ln\frac{\mu^2+T^2}{\Lambda^2}}\right)}
\def\ydm{\ln\frac{1}{x_m}}

\def\Qtmax{Q^2_{max}}
\def\del{\partial}

\def\Eq#1{Eq.~(\ref{#1})}
\def\bg{\begin{eqnarray}}
\def\nd{\end{eqnarray}}

\def\Fig#1{Fig.~\ref{#1}}

\begin{document}

\title{Energy dependence of jet transport parameter and
parton saturation in quark-gluon plasma}

\author{Jorge Casalderrey-Solana }
\affiliation{Nuclear Science Division, MS 70R0319, Lawrence Berkeley National
                   Laboratory, Berkeley, CA 94720}
\author{Xin-Nian Wang}
\affiliation{Nuclear Science Division, MS 70R0319, Lawrence Berkeley National
                   Laboratory, Berkeley, CA 94720}

\date{\today}

\begin{abstract}
We study the evolution and saturation of the gluon distribution
function in the quark-gluon plasma as probed by a propagating parton and its effect
on the computation of jet quenching or transport parameter $\hat{q}$. For thermal
partons, the saturation scale $Q^2_s$ is found to be proportional to
the Debye screening mass $\mu_D^2$. For hard probes, evolution at
small $x=Q^2_s/6ET$ leads to jet energy dependence of $\hat{q}$.
We study this dependence for both a conformal gauge theory in weak and strong
coupling limit and for (pure gluon) QCD. The energy dependence can
be used to extract the shear viscosity $\eta$ of the medium since $\eta$ can
be related to the transport parameter for thermal
partons in a transport description. We also derive
upper bounds on the transport parameter for both energetic and thermal partons.
The latter leads to a lower bound on shear viscosity-to-entropy density
ratio which is consistent with the conjectured lower bound $\eta/s\geq 1/4\pi$.
We also discuss the implications on the study of jet quenching at the 
BNL Relativistic Heavy Ion Collider and the CERN Large Hadron Collider and 
the bulk properties of the dense matter.
\end{abstract}

\maketitle

\section{Introduction}

Experimental data from the Relativistic Heavy-ion Collider (RHIC)
have shown significant
suppression of both high transverse momentum single inclusive
hadron spectra and back-to-back dihadron correlation in
central high-energy heavy-ion
collisions \cite{Adcox:2001jp,Adler:2002xw,starjet}. The
observed jet quenching phenomena can be attributed to parton
energy loss and medium  modification of the effective
parton fragmentation functions \cite{Gyulassy:2003mc,Wang:2004dn,Jacobs:2004qv}
due to gluon bremsstrahlung induced by multiple parton scattering.

Within the picture of multiple parton scattering in QCD, the
energy loss for an energetic parton propagating in a dense medium
is dominated by induced gluon bremsstrahlung. Taking into account
of the non-Abelian Landau-Pomeranchuck-Midgal (LPM) interference,
the radiative parton energy loss \cite{bdmps},
\begin{equation}
\Delta E= \frac{\alpha_s N_c}{4} \hat q_R L^2,
\label{bdmps}
\end{equation}
is found to depend quadratically on the medium length $L$ and a
jet transport or energy loss parameter,
\begin{equation}
\hat q_R=\rho \int dq_T^2 \frac{d\sigma_R}{dq_T^2} q_T^2,
\label{eq:qhat0}
\end{equation}
which describes the averaged transverse momentum transfer squared
per unit distance (or mean-free-path). Here $R$ is the color representation
of the propagating parton in $SU(3)$ and $\rho$ is the color charge density of
the medium. According to this picture, jet quenching as observed in
high-energy heavy-ion collisions is a direct measurement of
the jet transport parameter $\hat q_R$ in dense medium which
not only characterizes the color charge density but also the
interaction strength between the propagating parton and the medium.

Phenomenological studies based on variations of the parton energy
loss picture
\cite{Vitev:2002pf,Wang:2003mm,Eskola:2004cr,Turbide:2005fk} all
indicate the formation of an extremely high density matter in the
initial stage of high-energy heavy-ion collisions at the RHIC
energy. The averaged transport parameter extracted from different
phenomenological studies of the single inclusive high $p_T$ hadron
suppression in the most central $Au+Au$ collisions at RHIC is
\cite{majumder} $\hat q_F \sim 1 - 15$ GeV$^2$/fm (for a
propagating quark) at an initial time $\tau_0=1$ fm/$c$. A recent
simultaneous fit of the next-to-leading order (NLO) pQCD
calculation to both single and back-to-back dihadron suppression
\cite{Zhang:2007ja} narrows the uncertainty to $\hat q_F=1.1 -
1.4$ GeV$^2$/fm, which is still about 100 times higher than that
in a cold nucleus $\hat q_F\approx 0.013$ GeV$^2$/fm as extracted
from leading hadron suppression in deeply inelastic scattering off
large nuclei \cite{ww02}.

In most of the theoretical studies of parton energy
loss \cite{bdmps,Gyulassy:1993hr,Zakh,Wie,GLV}, except the
twist-expansion approach \cite{GuoW}, a static potential
model for jet interaction with the medium was assumed which
led to the factorized dependence of parton
energy loss on the transport parameter $\hat q_R$ in \Eq{bdmps}.
In this static potential model, energy and longitudinal
momentum transfer between a jet parton and the medium is ignored.
Therefore, elastic energy loss due to the recoil of the
medium parton during the jet-medium interaction is neglected
in the calculation of radiative parton energy loss. Furthermore,
the static potential model does not include the effect of
inelastic break-up (or parton radiation) of the medium partons
which can give rise to jet energy dependence of the transport
parameter $\hat q_R$. In a dynamical picture, the transport parameter
can be related to gluon distribution density
of the medium \cite{bdmps}. The jet energy dependence of the
transport parameter is then directly related to the scale and
momentum fraction dependence of the gluon distribution density.
Understanding the jet energy dependence
of the transport parameter not only helps us to improve the
phenomenological study of experimental data on jet quenching
but also provides additional information about the structure of the
dense quark-gluon matter in heavy-ion collisions. Furthermore,
as illustrated in a recent study \cite{Majumder:2007zh}, the
low energy limit ($E\sim T$ temperature of the medium) of the
transport parameter in jet quenching is directly related to the
shear viscosity of the quark-gluon matter in a transport description.
Therefore, experimental and theoretical study of the jet
energy dependence of the transport parameter will be able to
provide another piece of important information on bulk properties
of the dense medium.

Recently, the transport parameter $\hat q_R$ has also been calculated
for a strongly coupled $\mathcal{N}=4$ supersymmetric Yang-Mills (SYM) theory
in different limiting scenarios. With a definition in terms of an
adjoint Wilson loop along the light-cone, Liu, Rajagopal and
Wiedemann \cite{Liu:2006ug}
found that $\hat q_A$ in the large limit of the t' Hooft coupling
$\lambda=N_cg^2$ in SYM,
\begin{equation}
\hat q_A =\frac{\pi^{3/2}\Gamma(3/4)}{\Gamma(5/4)}\sqrt{\lambda}T^3,
\label{eq:ads1}
\end{equation}
scales with the temperature cubed and is independent of the
jet (propagating parton) energy. In another limit for a slowly
moving heavy quark, the transport parameter,
\begin{equation}
\label{kappaTADS}
\hat q_F =2\pi\sqrt{\lambda \gamma} T^3,
\end{equation}
as defined in Eq.~(\ref{eq:qhat0}), is found \cite{Gubser:2006nz,jorge:2007qw},
to depend on the square-root of the heavy-quark energy, where
$\gamma=E/M < (M/\sqrt{\lambda} T)^2$. It is not clear how these
two results are related to each other, though both describe the
transport properties in a SYM theory.

In this paper, we investigate the jet energy dependence of the
transport parameter $\hat q_R$ within perturbative QCD (pQCD).
We will first re-exam the relationship between the transport
parameter and the unintegrated gluon distribution function of the
color charges in the medium and how they are related to parton
energy loss in the medium. For energetic jet partons, there are
large logarithms of both momentum scale and small momentum
fraction. They allow us to take double logarithmic approximation
(DLA) and resum gluon radiation of the target color charges to
all orders. The initial condition to such a resummed evolution
of the gluon distribution can be calculated perturbatively
within pQCD at finite temperature with
hard thermal loop (HTL) resummation. From such resummed gluon
distribution one can further take into account gluon saturation
and calculate the saturation scale self-consistently which will
determine the transport parameter and its jet energy dependence.

\section{Parton energy loss, gluon distribution function and
transverse momentum broadening}

\begin{figure}
\centerline{\includegraphics[width=8cm]{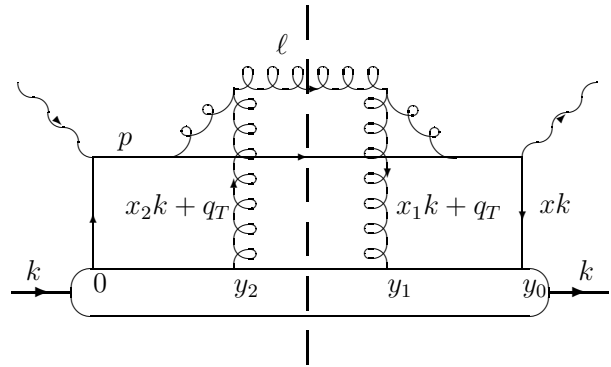}}
\caption{Feymann diagram for induced gluon radiation that
contibutes to the quark energy loss.}
\label{fig-diagram}
\end{figure}

Multiple
parton scattering within the high-twist expansion framework \cite{LQS}
can go beyond the static potential model and include energy and
longitudinal momentum transfer in the calculation of medium
modified fragmentation functions. The total energy loss
for a propagating parton in a deeply inelastic scattering (DIS)
off a large nucleus as shown in Fig.~\ref{fig-diagram} due to 
secondary quark-gluon scattering in this framework can be 
expressed as \cite{GuoW}
\begin{eqnarray}
\frac{\Delta E}{E}
&=& N_c\alpha_s \frac{2\alpha_s C_R}{N_c^2-1}
\int\frac{d^2q_T}{(2\pi)^2}\int d\ell_T^2 \int_0^1 dz \nonumber \\
&& \hspace{0.2in} \times \frac{1+(1-z)^2}{\ell_T^2(\ell_T^2+\mu_T^2)}
\frac{T_{qg}^A(x_B,x_L,q_T)}{f_q^A(x_B)}, \label{eq:dis0}
\end{eqnarray}
where
\begin{equation}
x_L=\frac{\ell_T^2}{2z(1-z)p\cdot k}, \,\,\,\,
\end{equation}
is the total longitudinal momentum transfer related to
induced gluon radiation with final transverse momentum $\ell_T$.
A similar longitudinal momentum transfer
\begin{equation}
x_T=\frac{q_T^2-2{\bf q}_T\cdot{\bf \ell}_T}{2(1-z) p\cdot k}\,\, ,
\end{equation}
is always provided by the initial gluon with transverse momentum $q_T$.
As illustrated in Fig.~\ref{fig-diagram}, $p=[0,p^-,0_T]$ is the 
initial quark momentum after its interaction with the photon, 
$k=[k^+,0,0_T]$ is the momentum per nucleon in the medium, 
$q_T$ is the transverse momentum of the gluon exchange with the 
medium, $\ell_T$ is the transverse momentum and $z$ is the fractional 
longitudinal momentum carried by the radiated gluon with four-momentum
$\ell=[\ell_T^2/2zp^-, zp^-,\vec \ell_T]$. The quark distribution
function $f_q^A(x_B)$ represents the production rate of the
initial quark carrying $x_B$ fraction of the nucleon longitudinal
momentum in DIS. Eq.~(\ref{eq:dis0}) is derived for quark
energy loss and one can extend it for gluon by replacing the
corresponding Casimir factor $C_R$ for gluons. In the collinear
expansion of the twist expansion approach, one normally makes
Taylor expansion of the hard partonic parts in $q_T$ and only the
quadratic terms lead to the twist-four contribution. One can,
however, approximate higher twist contributions from the
$q_T$-dependence of the hard partonic part of the multiple
scattering by using the average value $\langle
q_T^2\rangle=\mu_T^2$ in the cross section. As an extension of the
twist expansion, we will keep the integration over the gluon's
transverse momentum $q_T$. The unintegrated quark-gluon correlation
function is defined as,
\begin{eqnarray}
T^A_{qg}(x,x_L,q_T)&=& \int
\frac{dy^-_0}{2\pi}\, dy_1^-dy_2^- d^2\xi_T
 e^{i(x+x_L)k^+y^-_0} \nonumber \\
&&\hspace{-0.5in}\times (1-e^{-ix_Lk^+y_2^-})(1-e^{-ix_Lk^+(y^-_0-y_1^-)}) \nonumber \\
&&\hspace{-0.5in} \times e^{ix_Tk^+\xi^- - i{\bf q}_T\cdot{\bf \xi}_T}
\theta(-y_2^-)\theta(y^-_0-y_1^-) \nonumber \\
&& \hspace{-0.9in}\times\langle A | \bar{\psi}_q(0)\,
\frac{\gamma^+}{2}F_{\sigma}^{\ +}(y_{2}^{-})\, F^{+\sigma}(y_1^{-})\,\psi_q(y^-_0)
| A\rangle  ,
\label{eq:qgmatrix}
\end{eqnarray}
where $\xi=y_1 - y_2$, $y_0$, $y_1$ and $y_2$ are space-time coordinates
associated with the quark and gluon fields as illustrated 
in Fig.~\ref{fig-diagram}. The relative transverse coordinate
$\xi_T$ is the Fourier conjugate of the transverse momentum $q_T$
in the gluon distribution function. 

Even though the above parton energy loss is derived for quark
production and propagation in DIS, it is also valid for high-energy
heavy-ion collisions. In the latter case, we assume the life-time of
the quark-gluon plasma to be much longer than its formation time
and expansion time scale and therefore can treat thermal partons
inside the produced dense matter as in
asymptotic states. One therefore can neglect correlation between
the initial production rate of the jet parton and the quark and gluon
density of the produced medium. The quark-gluon correlation function
will then take a factorized form,
\begin{eqnarray}
\frac{T^A_{qg}(x,x_L,q_T)}{f_q^A(x)}
&=&\int dy^- \int\frac{d^3k}{(2\pi)^3 2k^+} f(k,y)
d\xi^- d^2\xi_T \nonumber \\
&& \hspace{-0.8in} \times e^{ix_Tk^+\xi^--i{\bf q}_T\cdot{\bf \xi}_T}
\langle k | F_{\sigma}^{\ +}(y_{2}^{-})\, F^{+\sigma}(y_1^{-})
| k \rangle \nonumber \\
&& \hspace{-1.05in}\times\left[e^{ix_Lk^+\xi^-}(1-e^{ix_Lk^+y^-})
+c(x_L)(1-e^{-ix_Lk^+y^-})\right]
 \nonumber \\
&& \hspace{-0.9in} =\pi \int dy^- \int\frac{d^3k}{(2\pi)^3} f(k,y)
\left[1-\cos(x_Lk^+y^-)\right] \nonumber \\
&& \hspace{-0.7in}\times\left[\phi_k(x_T+x_L,q_T)
+ c(x_L) \phi_k(x_T,q_T) \right] \, , \label{eq:corr2}
\end{eqnarray}
where $f(k,y)$ is the local phase space distribution of the
color sources in the medium and $c(x_L)=f_q(x+x_L)/f_q(x)$ is
the relative initial quark distributions in DIS
and is given by the corresponding ratio of jet production cross
sections in heavy-ion collisions. The unintegrated gluon distribution 
function per color source $\phi_k(x,q_T)$ is defined as
\begin{eqnarray}
\phi_k(x,q_T) &=& \int \frac{d\xi^-}{2\pi k^+} d^2\xi_T\,
e^{ixk^+\xi^- - i{\bf q}_T\cdot {\bf \xi}_T} \nonumber \\
&&\hspace{0.3in}\times 
\langle k| F^{\sigma +}(0)F_{\sigma}^{+}(\xi^-,{\bf \xi}_T)|k\rangle\, .
\end{eqnarray}

The structure of the quark-gluon correlation function in Eq.~(\ref{eq:corr2})
corresponds to two different bremsstrahlung processes and their 
interference \cite{GuoW} associated with the different
pole structures in Fig.~\ref{fig-diagram}. 
One can also categorize them according to
how the longitudinal momentum transfer $x_L$ is provided. In the
first term, the final gluon is induced by the secondary scattering
with the medium gluon in which the intermediate gluon is off-shell. 
The longitudinal momentum transfer $x_L$ is 
therefore provided by the medium gluon and the contribution is
proportional to gluon distribution $\phi_k(x_T+x_L,q_T)$ per medium
or ``constituent'' parton. These secondary processes correspond to 
quark-gluon Compton scattering where the initial gluon comes from a 
thermal constituent parton with a gluon distribution $\phi_k(x_T+x_L,q_T)$.
Among these processes, one can identify a special case in which the 
quark scatters directly with a medium constituent gluon ($x_L=1$) as purely
elastic processes and the corresponding energy loss as the
conventional elastic energy loss \cite{Wang:2006qr}. The
second term in Eq.~(\ref{eq:corr2}) corresponds to the processes in
which gluon radiation is induced by the hard scattering that 
produces the initial jet (the quark after the photon interaction 
in Fig. ~\ref{fig-diagram} is off-shell) and the final gluon scatters 
again with a soft medium gluon with momentum fraction $x_T$. It is therefore
proportional to the soft gluon distribution $\phi_k(x_T,q_T)$. The
longitudinal momentum transfer $x_L$ of the bremsstrahlung in this 
case is provided by the initial hard process with the cross section 
given in $c(x_L)$.

One can now define a generalized jet transport parameter,
\begin{eqnarray}
\hat q_R(E,x_L,y)&=&\frac{4\pi^2\alpha_sC_R}{N_c^2-1}
\int\frac{d^3k}{(2\pi)^3} f(k,y) \nonumber \\
&&\hspace{0.0in} \times \int
\frac{d^2q_T}{(2\pi)^2} \phi_k(x_T+x_L,q_T),
\label{eq:qhat-phi}
\end{eqnarray}
which includes the extra longitudinal momentum transfer $x_L$ from the
medium to the propagating parton and the radiated gluon. The total 
parton energy loss from Eq.~(\ref{eq:dis0}) can be expressed as
\begin{eqnarray}
\frac{\Delta E}{E}=\frac{\alpha_s N_c}{\pi}
\int dy^-dz {d\ell_\perp^2}
\frac{1+(1-z)^2}{\ell_T^2(\ell_T^2+\mu^2_T)}
[ \hat q_R(E,x_L,y) \nonumber \\
+ c(x_L) \hat q_R(E,0,y) ]
\sin^2\left[\frac{\ell_T^2y^-}{4Ez(1-z)}\right],
\hspace{0.2in}
\label{eq:de-twist}
\end{eqnarray}
in terms of the generalized jet transport parameter.
The first term with the generalized
transport parameter involves energy transfer between the
propagating parton and the medium. It contains (but not limited to)
what is normally defined as pure elastic energy loss \cite{Wang:2006qr}.
The second term that is proportional to the normal (or special) transport
parameter $\hat q_R(E,y)=\hat q_R(E,x_L=0,y)$ corresponds to pure
radiative energy loss.

Completing the integration over the phase-space of the radiated
gluon, one can recover from the second term a similar form of total
radiative energy loss in a static and uniform medium with finite
length as in Eq.~(\ref{bdmps}). However, one needs to know 
the $x_L$-dependence of the unintegrated gluon distribution function
in order to calculate the ``elastic'' part of the energy loss.
Furthermore, the transport parameter as defined in
Eq.~(\ref{eq:qhat0}) should have some non-trivial jet energy ($E$)
and temperature ($T$) dependence.

Within the framework of twist expansion, the transverse momentum
broadening of the quark jet has also been calculated \cite{Guo:1998rd},
\begin{equation}
\langle \Delta p_T^2\rangle
=\frac{4\pi \alpha_s C_R}{N_c^2-1} \frac{T_{qg}^A(x,0)}{f_q^A(x)}
=\int dy^- \, \hat q_R(E,0,y),
\end{equation}
which is directly related to the normal transport parameter 
$q_R(E,y)\equiv q_R(E,0,y)$ as defined in Eq.~(\ref{eq:qhat-phi})
in terms of the unintegrated gluon distribution density of the medium.
The jet transport parameter $q_R(E,y)$ therefore can be interpreted 
as the transverse momentum broadening per unit length for the
propagating parton, as defined in Eq.~(\ref{eq:qhat0}). Resummation
of higher twist contributions leads to a diffusion equation for the 
transverse momentum distribution in which the above is the averaged
transverse momentum broadening \cite{Majumder:2007hx}.

The approach leading to the above total energy loss and transverse
momentum broadening has gone beyond the conventional static potential
model in two aspects: (a) The result includes the longitudinal momentum 
transfer ($x_T$) between the jet parton and the medium parton, which 
is related to the transverse momentum transfer ($q_T$) through the
unintegrated gluon distribution density of the medium. This will 
result in the jet energy dependence of both the generalized and the 
normal transport parameter $\hat q_R(E,y)$ (or transverse momentum broadening)
which is absent in the static potential model. 
(b) The formula also includes the processes in which longitudinal momentum 
transfers $x_L$ comes from the medium parton and therefore it depends 
on the generalized jet transport parameter $\hat q_R(x_L)$. 
It contains contributions from elastic energy loss.

For the remainder of this paper we will focus on the energy
dependence of the normal transport parameter $\hat q_R \equiv \hat q_R(E,y)$.
Since it is essentially the transverse momentum broadening per unit
length which can be directly measured in experiments such as DIS 
and $\gamma$-jet events in heavy-ion collisions, we will suppress 
the space and time dependence to simplify the notation.

\section{Gluon distribution in a quark-gluon plasma}

As shown in \Eq{eq:qhat-phi}, the transport parameter $\hat q_R$
experienced by a propagating parton
can be defined in terms of the unintegrated gluon distributions
$\phi_k(x,q_T^2)$ of the color sources in the quark-gluon plasma. After
averaging over the momentum of the color sources, it
can be expressed as,
\begin{eqnarray}
\hat q_R&=&\frac{4\pi^2C_R}{N_c^2-1} \rho \int_0^{\mu^2}
\frac{d^2q_T}{(2\pi)^2}\int dx
 \nonumber \\
&&\hspace{0.2in} \times \delta(x-\frac{q_T^2}{2p^-\langle k^+\rangle})
\alpha_s(q_T^2)\phi(x,q_T^2),
\label{eq:xg1}
\end{eqnarray}
where $\langle k^+\rangle$ is the average energy of the color sources
and $\phi(x,q_T^2)$ is the corresponding average unintegrated gluon
distribution function per color source.
The integrated gluon distribution is
\begin{equation}
xG(x,\mu^2)=\int_0^{\mu^2} \frac{d^2q_T}{(2\pi)^2}  \phi(x,q_T).
\end{equation}
We have extended our
earlier definition of $\hat q_R$ to include the case of
a running strong coupling constant $\alpha_s$ in QCD.
We will refer to the case of fixed coupling constant as conformal
gauge theory. However, for
any scale below the temperature $\mu^2\leq T^2$ we will
consider $\alpha_s$ frozen and treat it as a constant.

Consider the lowest order parton-parton small angle scattering,
we can obtain $\hat q_R$ as,
\begin{eqnarray}
\hat q_R&=&\sum_b \nu_b  g^4C_{Rb}
\int \frac{d^3k}{(2\pi)^3} f_b(k)(1\pm f_b(k^\prime))
q^2_T|{\cal M}_{Rb}|^2
\nonumber \\
&&\times\frac{d^3k^\prime}{(2\pi)^3}
\frac{d^3p^\prime}{(2\pi)^3}\,
(2\pi)^4\delta^4(p+k-p^\prime-k^\prime),
\label{eq:qhat1}
\end{eqnarray}
where ${\cal M}_{Rb}$ is the truncated parton-parton scattering
matrix element,
\begin{eqnarray}
{\cal M}_{Rb}\approx \left[ \frac{1}{q^2+\mu_D^2\pi_L(x_q)} \right.
&& \nonumber \\
&&\hspace{-1.2in}\left.
-\frac{(1-x_q^2)\cos\phi}{q^2(1-x^2_q)+\mu_D^2\pi_T(x_q)+\mu_{\rm mag}^2}\right]\, ,
\end{eqnarray}
where $\cos\phi=(\vec{v}\times\vec{q})\cdot(\vec{v}_b\times\vec{q})/q^2$
, $x_q=q_0/q$ and $\mu_D^2=g^2(N_c+n_f/2)T^2/3$ is the Debye screening
mass in thermal QCD medium with temperature $T$
and $\mu_{\rm mag}\approx N_cg^2/2\pi$ is the non-perturbative magnetic
screening mass \cite{Biro:1992wg,Alexanian:1995rp,vonHippel:2002ih} .
The color factors for different scatterings are $C_{qq}=C_F/2N_c$,
$C_{qg}=1/2$, $C_{gg}=N_c^2/(N_c^2-1)$.
The statistical factor $\nu_b$ is $2(N_c^2-1)$ for
gluons and $4N_cn_f$ for $n_f$ flavors of quarks.
We use an effective gluon propagator
to include the resummation of hard thermal loops (HTL) \cite{Pisarski:1988vd}.
The scaled self-energies in the effective propagator
in the long-wavelength limit are given by \cite{Weldon:1982aq},
\begin{eqnarray}
  \pi_L(x_q)&=&1-\frac{x_q}{2}
  \ln\left(\frac{1+x_q}{1-x_q}\right) + i \frac{\pi}{2}x_q
  \, , \label{eq:self1}\\
  \pi_T(x_q)&=&\frac{x^2_q}{2}+\frac{x_q}{4}(1-x^2_q)
  \ln\left(\frac{1+x_q}{1-x_q}\right) \nonumber \\
&& - i \frac{\pi}{4}x_q(1-x^2_q)
  \, .\label{eq:self2}
\end{eqnarray}
One can rewrite the phase-space integration in Eq~(\ref{eq:qhat1}) as
\begin{eqnarray}
&&\int \frac{d^3k^\prime}{(2\pi)^3}
\frac{d^3p^\prime}{(2\pi)^3}\,
(2\pi)^4\delta^4(p+k-p^\prime-k^\prime) \nonumber \\
&&\hspace{0.5in}
=\frac{1}{(2\pi)^2}\int dx d^2q_T\delta (x-\frac{q_T^2}{2p^-k^+}),
\end{eqnarray}
where $x=q^+/k^+$. For small angle scattering, one can set
$q^2\approx q^2_T$ and $x_q\approx x\,k^+/q_T$. We further
approximate $k^+$ by its average value $\langle k^+\rangle=3T$ in
the scattering matrix. Note that energy-momentum conservation
fixes the relative angle between $k$ and $q$. Therefore, the
angular phase-space for $k$ is only $2\pi$. One can complete the
rest of the phase-space integration over the initial momentum,
\begin{eqnarray}
\int \frac{k^2dk}{4\pi^2}f_b(k)(1\pm f_b(k^\prime))
&\approx& \int \frac{k^2dk}{4\pi^2}f_b(k)(1\pm f_b(k))
\nonumber \\
&&\hspace{-1.0in} =\frac{T^3}{12}\,\, {\rm (gluons)\,\,\, or}\,\,
\frac{T^3}{24}\,\, ({\rm quarks}).
\end{eqnarray}
Using
\begin{equation}
\frac{1}{2}C_{Rq}\nu_q+C_{Rg}\nu_g
=2N_cC_R(1+\frac{n_f}{2N_c})\, ,
\end{equation}
and
\begin{equation}
\rho=\frac{T^3}{\pi^2}\zeta(3)(\nu_g+\frac{3\nu_q}{4})
=2(N_c^2-1)(1+\frac{3n_f}{4C_F})\frac{T^3}{\pi^2}\zeta(3),
\end{equation}
one can express Eq.~(\ref{eq:qhat1}) as
\begin{eqnarray}
\hat q_R&=&\frac{4\pi^2\alpha_sC_R}{N_c^2-1}\rho N_c\frac{\alpha_s}{2\pi}
\frac{\pi^2}{6\zeta(3)}\frac{1+n_f/2N_c}{1+3n_f/4C_F} \nonumber \\
&\times& \int dx dq_T^2 \delta(x-\frac{q_T^2}{2p^-\langle k^+\rangle})
q_T^2 |{\cal M}_{Rb}|^2 \, .
\label{eq:qhat2}
\end{eqnarray}
The factor $\pi^2/6\zeta(3)$ comes from the quantum
statistics effect for the final state partons in the
scattering processes.
According to the definition in Eq.~(\ref{eq:xg1}), one can obtain
the unintegrated gluon distribution function,
\begin{equation}
\phi(x,q_T^2)=2N_c\alpha_s
\frac{\pi^2}{6\zeta(3)}\frac{1+n_f/2N_c}{1+3n_f/4C_F}
|{\cal M}_{Rb}|^2 q_T^2\, ,
\label{eq:unintg0}
\end{equation}
in a quark-gluon plasma and the integrated gluon distribution function is
\begin{eqnarray}
xG(x,\mu^2)&=&\frac{N_c\alpha_s}{2\pi}
\frac{\pi^2}{6\zeta(3)}\frac{1+n_f/2N_c}{1+3n_f/4C_F} \nonumber \\
& & \times \int_0^{\mu^2} dq_T^2 |{\cal M}_{Rb}|^2 q_T^2\, .
\label{eq:xg-qgp1}
\end{eqnarray}

We concentrate in the small $x_q$ region, $x_q=3xT/q_T \ll 1$.
For a typical momentum transfer of the order of $\mu_D$, the requirement
that $x_q$ is small leads to $x\ll (\sqrt{N_cg^2/3})/3$. Within this
approximation, we obtain
\begin{eqnarray}
\pi_L(x_q) \approx 1-ix \frac{3\pi T}{2 q_T}, \,\,\,\,\,\,
\pi_T(x_q) \approx -ix \frac{3\pi T}{4 q_T},
\end{eqnarray}
and
\begin{eqnarray}
\phi(x,q_T^2)&=&\frac{2N_c\alpha_s}{\mu_D^2}
\frac{\pi^2}{6\zeta(3)}\frac{1+n_f/2N_c}{1+3n_f/4C_F} \widetilde \phi(x,y_q);
\label{eq:unintg1} \\
\widetilde \phi(x,y_q)&\equiv &\mu_D^2 |{\cal M}_{Rb}|^2 q_T^2  \nonumber \\
&\approx&
\frac{y^2_q}{y_q(y_q+1)^2+x^2 9\pi^2T^2/4\mu_D^2}  \label{eq:unintg} \\
&+& \frac{1}{2}\frac{y^2_q}{y_q(y_q+\mu_{\rm mag}^2/\mu_D^2)^2+x^2 9\pi^2T^2/16\mu_D^2},\nonumber
\end{eqnarray}
where $y_q=q_T^2/\mu_D^2$. For $x\gg 4\mu_{\rm mag}/\pi =2N_cg^2/\pi^2$, one can
neglect the magnetic mass and complete the integration
in Eq.~(\ref{eq:xg-qgp1}) and obtain
\begin{eqnarray}
xG(x,\mu^2)&\approx&N_c\frac{\alpha_s}{2\pi}
\frac{\pi^2}{6\zeta(3)}\frac{1+n_f/2N_c}{1+3n_f/4C_F} \nonumber \\
&\times& \left\{\left[\ln(1+\frac{\mu^2}{\mu_D^2})
-\frac{\mu^2/\mu_D^2}{1+\mu^2/\mu_D^2}\right]   \right. \nonumber \\
&\times&\left[1-0.035\frac{3xT}{\mu_D}\right]
e^{-3xT\mu_D/\mu^2} \nonumber \\
&+&\left. \frac{1}{6}\ln(1
+\frac{16}{9\pi^2}\frac{\mu^6}{x^2T^2\mu_D^4})\right\} \, ,
\end{eqnarray}
where the first term is an approximation of the numerical
integration from the electric part of the interaction for $x\le
2\mu_D/3\pi T = 2\sqrt{N_cg^2/3}/3\pi$. Because of the static
Debye screening, it has a very weak $x$ dependence in this $x$
region, which can be ignored for large values of $\mu/\mu_D \ge
1$. 
 The magnetic
part of interaction, on the other hand, has only dynamical
screening and therefore lead to the dominant $x$ dependence of the
gluon distribution from a quark-gluon plasma at small $x$.
However, our approximations are not valid for 
$x\le 4\mu_{\rm mag}/3\pi T \approx 2N_cg^2/3\pi^2$,
where the
the non-perturbative magnetic mass
\cite{Biro:1992wg,Alexanian:1995rp,vonHippel:2002ih} $\mu_{\rm
mag}\approx N_cg^2/2\pi$ becomes important. In this region, the
logarithmic $x$-dependence of the gluon distribution from the
magnetic interaction disappears and is replaced by a constant
$\ln(\mu^2/\mu_{\rm mag}^2)$.

For large $\mu^2/\mu_D^2\gg 1$ in the small $N_cg^2 < x < \sqrt{N_c}\, g$
region of a pure gluonic plasma ($n_f=0$), the gluon distribution
per gluonic color source is then,
\begin{equation}
xG(x,\mu^2)\approx C_A \frac{\alpha_s}{\pi}
\frac{\pi^2}{6\zeta(3)}\frac{1}{2}
\left[\frac{3}{2}\ln\frac{\mu^2}{\mu_D^2}
+\frac{1}{3}\ln\frac{\mu_D}{xT}\right] \, ,
\label{eq:g-g1}
\end{equation}
which is generated from perturbative gluon radiation.
For a pure quark plasma, the corresponding gluon distribution
for each quark color source is
\begin{equation}
xG(x,\mu^2)\approx C_F \frac{\alpha_s}{\pi}
\frac{\pi^2}{6\zeta(3)}\frac{1}{3}
\left[\frac{3}{2}\ln\frac{\mu^2}{\mu_D^2}
+\frac{1}{3}\ln\frac{\mu_D}{xT}\right] \, .
\label{eq:g-q1}
\end{equation}
For the remainder of this paper, we will focus on a pure gluonic plasma.

\section{\label{satHTL}Gluon saturation in a plasma}

\begin{figure}
\centerline{\includegraphics[width=9cm]{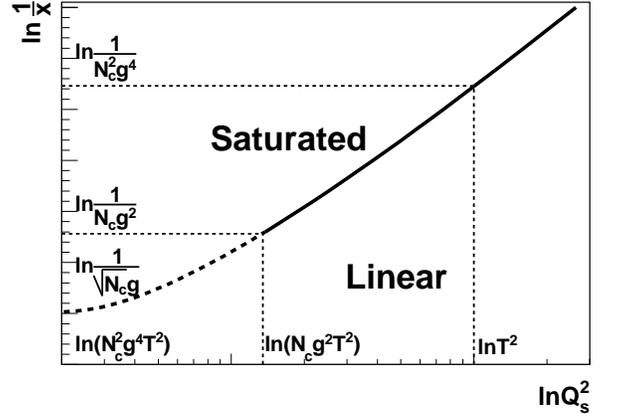}}
\caption{Illustration of the hierarchy of the saturation scale $Q_s^2(x)$
below the hard scale $\mu^2=T^2$ in a weak coupling gluonic plasma.}
\label{fig-qs}
\end{figure}

Similar to gluon saturation in a large nucleus at small $x$,
saturation could also happen in the small $x$ region of a
quark-gluon plasma. The saturation scale is given by \cite{Mueller:1999wm}
\begin{equation}
\label{Qsdef}
Q_s^2(x)=\frac{4\pi^2 N_c \alpha_s}{N_c^2-1}\rho xG(x,Q_s^2) \min(L,L_c)
\end{equation}
where $L_c=1/xT$ is the coherence length for parton scattering in
a thermal medium. Since the HTL resummation does not include
coherence effects, the use of  the gluon distribution in
Eq.~(\ref{eq:g-g1}) requires that the mean-free-path of thermal
gluons must be larger than the coherence length. Given the
perturbative expression of the mean-free-path
\cite{Majumder:2007zh},
\begin{equation}
\lambda^{-1}_{f}=\langle\rho \sigma_{\rm tr}\rangle
\approx \frac{4\zeta(3)}{9\pi} N_c^2\alpha_s^2 T \ln\frac{1}{N_c\alpha_s},
\end{equation}
this implies,
\begin{equation}
\frac{L_c}{\lambda_{f}}=\frac{4\zeta(3)}{9\pi}
\frac{N_c^2\alpha_s^2}{x} \ln\frac{1}{N_c\alpha_s} \le 1,
\end{equation}
or
\begin{equation}
x\ge \frac{4\zeta(3)}{9\pi}
N_c^2\alpha_s^2 \ln\frac{1}{N_c\alpha_s}
\sim (N_c\alpha_s)^2\ln\frac{1}{N_c\alpha_s}.
\end{equation}
In this regime, one can  use the perturbative gluon
distribution [Eq.~(\ref{eq:g-g1})] to determine the saturation
scale,
\begin{eqnarray}
\label{fullQsT}
Q_s^2(x)&=&\frac{4\pi^2 N_c \alpha_s}{N_c^2-1}\rho xG(x,Q_s^2) L_c
\nonumber \\
&=&\frac{\pi}{x}(N_c\alpha_s)^2T^2
\left[\ln\frac{Q_s^2}{\mu_D^2}
+\frac{2}{9}\ln\frac{\mu_D}{xT}\right]\, .
\label{eq:htlsat}
\end{eqnarray}
Neglecting the logarithmic terms, one can get a simple expression for
the saturation scale in the perturbative regime,
\begin{equation}
\label{QsHTL}
Q_s^2(x)/T^2\sim \frac{(N_cg^2)^2}{x}.
\end{equation}
Since $\mu_D^2=N_c g^2 T^2/3 \sim N_cg^2T^2$, we note the following
hierarchy of the saturation scale in a perturbative gluonic plasma
\begin{eqnarray}
Q_s^2(x) &\sim& (N_c g^2)^2 T^2\sim \mu_{\rm mag}^2, {\rm for}\,\,\, x\sim 1
\nonumber \\
Q_s^2(x) &\sim& N_cg^2 T^2 \sim \mu_D^2,
{\rm for}\,\,\, x\sim N_cg^2\sim \frac{\mu_D^2}{T^2}\nonumber \\
Q_s^2(x) &\sim& T^2,{\rm for}\,\,\, x\sim  N_c^2g^4 \sim \frac{\mu_{\rm mag}^2}{T^2},
\end{eqnarray}
as illustrated in Fig.~\ref{fig-qs}.

In the calculation of $\hat q_R$ for interaction among thermal partons,
the typical $x_m=Q_s^2/\langle \hat s\rangle=Q_s^2/18T^2$. One can
determine the saturation scale at $x_m$ from
\Eq{eq:htlsat}
\footnote{Note that the typical $x_m\sim N_c g^2$; thus 
the breakdown of the approximation in \Eq{eq:g-g1} due to 
magnetic mass coincides with 
the onset of non-linear effects },
\begin{equation}
Q_s^2(x_m)\approx \mu_D^2\frac{3}{2}\sqrt{\frac{1}{\pi}\ln\frac{18T}{\mu_D}} \, .
\label{eq:qsth}
\end{equation}
It is interesting to note that the gluon saturation scale for interaction
among thermal partons coincides approximately with the Debye screening mass.
Therefore, resummation of HTL effectively provides some kind of mechanism
for gluon saturation in a thermal gluon plasma.

To obtain the transport parameter $\hat q_R$ at scale $\mu^2\le T^2$
due to interaction with the gluonic color sources via exchange of 
HTL gluons, one has to complete the integral in \Eq{eq:qhat2}. 
A simple integration in \Eq{eq:qhat2} without considering effect
of gluon saturation gives
\begin{equation}
\hat q_R\approx \frac{14}{15}
\pi N_c^2\alpha_s^2 T^3 \ln\frac{\mu^2}{\mu_D^2}\, ,
\label{eq:qhatth0}
\end{equation}
as obtained by a previous calculation of $\hat q_R$ with dynamic
screening \cite{Wang:2000uj}. One can also obtain the above result 
from the integrated gluon distribution
\begin{equation}
\hat q_R\simeq \frac{4\pi^2C_R}{N_c^2-1} \rho
\left[xG(x,\mu^2)\right]_{x=\mu_D^2/\mu^2} \, .
\label{eq:qhatth1}
\end{equation}

In principle, one should take into account the
effect of gluon saturation in evaluating the transport
parameter in the region $q_T^2<Q_s^2(x)$. In this regime, we
can follow KNL model \cite{Kharzeev:2002ei} and assume the saturated 
unintegrated gluon distribution as a constant in $q_T$,
\begin{eqnarray}
\phi(x,q_T^2)=\frac{2N_c\alpha_s}{\mu_D^2}\frac{\pi^2}{6\zeta(3)}
\left\{ \begin{array}{ll}
 \widetilde \phi(x,Q_s^2/\mu_D^2)
\, ,&
\,\, q^2_T<Q^2_s\, ; \\
& \\
 \widetilde \phi(x,q_T^2/\mu_D^2)\, , &  \,\,q^2_T>Q^2_s\, ,
\end{array}
\right. \,
\end{eqnarray}
where $\widetilde \phi(x,y_q)$ is given by Eq.~(\ref{eq:unintg}) and
the saturation scale $Q_s^2(x)$ is determined by Eq.~(\ref{eq:htlsat}).
Using the above model for gluon distribution in the saturated regime
in Eq.~(\ref{eq:xg1}), one can evaluate the thermal parton transport
parameter. The result,
\begin{equation}
\hat q_R\approx
\pi N_c^2\alpha_s^2 T^3 \ln\frac{Q_m^2}{\mu_D^2}\, ,
\label{eq:qhatth02}
\end{equation}
with $Q_m^2=18 T^2$, is nearly identical to Eq.~(\ref{eq:qhatth0}). 
This is because the dominant contribution to the parton transport 
parameter comes from $q_T^2>Q_s^2$ for large $Q_m^2/\mu_D^2>1$ and
therefore the effect of gluon saturation is negligible in the
calculation of transport parameter for thermal partons.

The similarity between results in Eqs.~(\ref{eq:qhatth02}) 
and (\ref{eq:qhatth0}) is also an indication that
saturation effect is already present in the unintegrated
gluon distribution function $\phi(x,q_T^2)$ in Eqs.~(\ref{eq:unintg1})
and (\ref{eq:unintg}) due to HTL resummation.  One can clearly
see this by analyzing the unintegrated gluon distribution
$\phi(x,q_T^2)$ [Eqs.~(\ref{eq:unintg1}) and (\ref{eq:unintg})]
at $x=q_T^2/\langle \hat s\rangle$. For
large $q_T^2\gg \mu_D^2$, $\phi(x,q_T^2)\sim 1/q_T^2$. The electric
contribution to $\phi(x,q_T^2)(x=q_T^2/\langle \hat s\rangle)$ reaches its
peak value $\sim N_c\alpha_s/\mu_D^2$
at $q_T^2\approx \mu_D^2\sim Q_s^2$ and vanishes at $q_T^2=0$. Without
the magnetic mass, the magnetic contribution
to $\phi(x,q_T^2)(x=q_T^2/\langle \hat s\rangle)$, however, continues to
increase at $q_T^2<Q_s^2\sim \mu_D^2$ and reaches a finite value
$\phi(x,q_T^2)(x=q_T^2/\langle \hat s\rangle)\sim 1/\mu_D^2$
at $q_T^2\ll \mu_D^4/T^2$. However, contribution to $\hat q_R$ from 
this region of limited phase space is sub-leading in the leading 
logarithmic approximation.

\section{\label{ETGD} Evolution of the Thermal Gluon Distribution}

The gluon distribution function in \Eq{eq:g-g1} was obtained via
parton interaction in a thermal medium with a HTL resummed gluon
propagator and, thus, is only valid for scales
$\mu^2 < T^2$ . At larger scales, radiation of hard modes, {\em \tie},
partons with momentum $k>T$, is possible. These processes lead to
the evolution of the gluon distribution which in vacuum
is governed by the BFKL/DGLAP equations in the linearized regime.
In the medium this evolution may be modified due to the interaction
of the radiated gluons with thermal partons. However,
since the medium effects
are of the order of $\mu_D<< T$, we neglect those at hard scales and
use the vacuum evolution to determine the gluon distribution.
The previous computation in \Eq{eq:g-g1} serves as an initial
condition of this evolution at $\mu^2=T^2$.

Since we are interested in the determination of $\hat{q}_R$ at
large jet energies, we need to know the unintegrated parton
distribution $\phi(x,q^2_T)$ in \Eq{eq:xg1} at small $x\sim
\left<q^2_T\right>/6 E T$. For a large path length, the typical
total momentum transfer, $\hat{q} L$, which will set the scale of
the process, is also large. This is the regime of the double
logarithmic approximation (DLA), in which the BFKL and DGLAP
equations coincide \cite{Gribov:1981ac,Mueller:1985wy}. In this
approximation, all terms enhanced by two large logarithms of the
type \be \label{DLAlogs}
\left(\alpha_s(k^2)N_c\ln\frac{k^2}{\mu^2}\ln\frac{1}{x}\right)^n
\,, \ee are resummed. Thus, the DLA approximation is valid if terms
of the above type are larger than those of type \be
\label{otherlogs} \left(\alpha_s(k^2)N_c\ln\frac{1}{x}\right)^n ,
\,\,\,\,\, \left(\alpha_s(k^2)N_c\ln\frac{k^2}{\mu^2}\right)^n \,.
\ee The resummation of the terms in \Eq{DLAlogs} leads to  the
evolution equation, \be \label{DLAeq} \frac{\del^2 xG(y,\xi)}{\del
y \, \del \xi}=\frac{1}{2}xG(y,\xi) \, , \ee where, following
Ref.~\cite{Gribov:1981ac}, we have defined variables $y$ and $\xi$
as \footnote{The definition we use is slightly different from that
of \cite{Gribov:1981ac} and is more suitable for the description
of a conformal plasma ($\alpha_s$ fixed) } \be \label{xidef}
\xi&=&\int^{Q^2}_{\mu^2} \frac{dk^2}{k^2} \frac{2\alpha_s(k)N_c}{\pi} \, ,\\
  y&=&  \ln\frac{1}{x} \, .
\ee The asymptotic solution to \Eq{DLAeq} leads to a growth of the
gluon distribution function of the order $\exp(\sqrt{2\xi y})$,
while resummation of terms in \Eq{otherlogs} leads to
$\exp(\alpha_s N_c y)$ and $\exp(\xi)$  respectively
\cite{Gribov:1981ac}. Therefore, the DLA approximation is valid as
long as \be \label{DGLAPc}
\xi &<<& \sqrt{\xi y}\,, \\
\label{BFKLc}
\alpha N_c y&<<& \sqrt{\xi y} \,.
\ee

Note that the definition of $\xi$ allows to describe
simultaneously the evolution of a conformal
and non-conformal theory. For these two cases we have
\be
\label{xiofQ}
\xi(Q^2)     =
\left\{ \begin{array}{ll}
 \blambda  \ln (Q^2/\mu^2) &\mbox{\,  \,\, for fixed\,}  \alpha_s \\
& \\
\frac{2 N_c}{\pi b} \ln \frac{\ln (Q^2/\Lambda^2)}
{\ln (\mu^2/\Lambda^2)} & \mbox{\,  \,\, for running\,} \alpha_s
\end{array}
\right. \,,
\ee
where the reduced t'Hooft coupling is $ \blambda =  2 N_c \alpha_s /\pi$ and
$b=(11N_c-2N_f)/12\pi$.

The general solution of \Eq{DLAeq} can be found by performing a
Laplace transformation
and is given by \cite{Mueller:1985wy,Dokshitzer:1977sg}
\be
\label{DLAgensol}
xG(x,Q^2)=\int^{a+i\infty}_{a-i\infty} \frac{dn}{2\pi i}
e^{ny+\frac{\xi}{2n}} D(n) \, ,
\ee
where $a$ is any real number larger than the real part of any poles of $D(n)$.
The corresponding Laplace transformation $D(n)$ of the gluon distribution in
\Eq{eq:g-g1} at $Q^2=\mu^2=T^2$ ($\xi=0$) is:
\begin{eqnarray}
\label{Dpert}
D(n)&=&\int^{\infty}_{0^-} dy e^{-n y} xG(x,T^2)
\nonumber \\
&=&\frac{N_c\alpha_s^T}{2\pi}
\frac{\pi^2}{6\zeta(3)}\frac{1}{3}
\left[\frac{1}{n} 4\ln\frac{T^2}{\mu^2_D}+\frac{1}{n^2}\right] \, ,
\end{eqnarray}
where $\alpha_s^T$ is the strong coupling constant $\alpha_s$ evaluated at
a scale that is proportional to $T^2$, since \Eq{eq:g-g1} is obtained
through scattering between thermal partons.

For large $y \xi$ values, the integral in \Eq{DLAgensol} can be performed
by saddle point approximation, yielding
\be
\label{xGevolved}
xG(x,Q^2)&=&
\frac{N_c\alpha_s^T}{2\pi}
\frac{\pi^2}{6\zeta(3)}\frac{1}{3}
\frac{e^{\sqrt{2\xi y}}}{\sqrt{\pi}(2\xi y)^{1/4}}
\nonumber \\
&& \hspace{0.2in}\times \left[
2\ln\frac{T^2}{\mu^2_D} + \frac{y}{(2\xi y)^{1/2}}
\right] \,.
\ee

The above evolved gluon distribution function grows rapidly
(faster than a power) with the rapidity $y$. Thus, at large $y$
non-linear effects become important leading to parton saturation.
Similarly, we can determine the saturation scale
$Q^2_s(x)$ from \Eq{Qsdef} with the above gluon distribution
function $xG(x,Q)$,
\be
\label{QsdefDLA}
Q^2_s=B(x,Q^2_s) \min(L,L_c)
    \exp\left\{\sqrt{2\xisat y}\right\} \, ,
\ee
where $\xisat=\xi(Q^2_s)$ and
\be
\label{Bdef}
B(x,Q^2_s)&=&\frac{1}{9}
\frac{\pi^3}{\zeta(3)}\frac{ N_c \alpha_s(Q^2_s) }{N_c^2-1}
 \rho
 \frac{ N_c \alpha_s^T}{\sqrt{ \pi}\left(2\xisat y\right)^{1/4}}
\nonumber \\
&&\times
\left[
2\ln\frac{T^2}{\mu^2_D} + \frac{y}{\left(2\xisat y\right)^{1/2}}
\right] \,.
\ee
In solving the self-consistent equation \Eq{QsdefDLA} we will
neglect the weak dependence of $B(x,Q_s)$ on $x$ and $\xisat$
and treat it as a constant as compared to the dependence in the
exponent. This is an approximation we will take throughout
this paper.

We now can use the evolved gluon distribution function in \Eq{xGevolved}
to compute the jet transport parameter as defined in \Eq{eq:xg1}. In the
linear evolution region ($q_T^2>Q^2_s$), the unintegrated parton
distribution is computed by taking the derivative of \Eq{xGevolved}
with respect to the scale.  Keeping the leading term
in $\xi y$ ({\em i.e.} consider only the $\xi y$ dependence in the exponent)
we find
\be
\label{phileading}
\phi_{DLA}(x,q^2_T)&=&4\pi \frac{\del}{\del q^2_T} xG(x,q^2_T)
\nonumber \\
&\approx&
8\frac{y}{\sqrt{2\xi y}}\frac{\alpha_s(q^2_T)N_c}{q^2_T} xG(x,q^2_T)\,.
\ee
Using \Eq{Qsdef}, we find at $q^2_T=Q^2_s$,
\be
\phi_{DLA}(x,Q^2_s)=\frac{2}{\pi^2}\frac{N^2_c-1}{\rho \min(L,L_c)}\frac{y}{\sqrt{2\xisat y}}
\, .
\ee

At scales $q^2_T< Q^2_s$, \Eq{xGevolved} is no longer valid since
saturation effects take place which tame the growth of the gluon
distribution function. Inspired by the KLN model of saturation
\cite{Kharzeev:2002ei}, we use a simplified model for the
unintegrated gluon distribution function \be \label{KLNphi}
\phi(x,q^2_T)  = \left\{ \begin{array}{ll}
\frac{2}{\pi^2}\frac{N^2_c-1}{\rho
\min(L,L_c)}\frac{y}{\sqrt{2\xisat y}}\, ,&
\,\, q^2_T<Q^2_s\, ; \\
& \\
\phi_{DLA}(x,q^2_T)\, , &  \,\,q^2_T>Q^2_s\, .
\end{array}
\right. \,
\ee
We can then express $\hat{q}_R$ in \Eq{eq:xg1} as
\be
\hat q_R&=&\frac{C_R}{N_c}\frac{4\pi^2 \rho}{N_c^2-1}
\int dx\left[\int_0^{Q^2_s}
\frac{d^2q_T}{(2\pi)^2}
\delta(x-\frac{q_T^2}{Q^2_{max}})  \right.
\nonumber\\
&&\times \alpha_s(q^2_T) N_c\phi_{DLA}(x,Q^2_s)
+\int_{Q^2_s}^{Q^2_{max}}
\frac{d^2q_T}{(2\pi)^2}
\nonumber\\
&&\times\left.
\delta(x-\frac{q_T^2}{Q^2_{max}}) \alpha_s(q^2_T) N_c \phi_{DLA}(x,q^2_T)
\right],
\ee
where  $Q^2_{max}\approx 6 E T$.
Integrating out the $\delta$-function, we have
\be
\label{qhatofx}
\hat{q}_R&=&\frac{C_R}{N_c}\frac{2}{\pi}\Qtmax\int^{x_m}_0 dx
\frac{\alpha_s(x \Qtmax) N_c}{\min(L,L_c)}\nonumber \\
&\times&\hspace{-0.1in}\frac{\yd}{\sqrt{2\yd\xi(x\Qtmax)}} +
\frac{C_R}{N_c}\frac{4\pi^2 \rho}{N_c^2-1} \nonumber \\
&\times&\hspace{-0.1in}\int^1_{x_m} dx
 \, N_c \alpha_s(x\Qtmax)  \phi_{DLA}(x,x\Qtmax),
\ee
where $x_m=Q^2_s/\Qtmax$.

\section{Conformal Plasma}

We first examine the behavior of the saturation scale and jet
transport parameter in a medium with fixed coupling constant. For
a medium length $L$ that is always larger than the coherence
length for any jet energy, we find \be \ln\frac{Q^2_s}{\mu^2}\sim
\ln\frac{1}{x} \, , \ee for small $x$. This means that both
constraints in Eqs.~(\ref{DGLAPc}) and (\ref{BFKLc}) are fulfilled at small
coupling $\blambda$. We can then use the DLA approximation to
describe the evolution of the gluon distribution function and
evaluate the saturation scale and transport parameter at small
$x_m\sim 1/E$. Note that the eikonal approximation is valid for
distances such that the total momentum transferred to the probe
$\hat{q}L\ll\Qtmax$, since $\Qtmax$ is the momentum transfer for a
large angle ($90^o$) scattering. From this requirement and
\Eq{qhatofxs} we find
\be
\label{llimit}
\bar \lambda L\ll L_c\frac{\Qtmax}{Q^2_s}\, ,
\ee
which is compatible with the weak coupling approximation and
$L>L_c$ if $\Qtmax\gg Q^2_s$.

We determine the saturation scale by solving the self-consistent
equation \Eq{QsdefDLA}. Treating $B$ as a constant and using
the definition of $\xi$ for fixed coupling constant [\Eq{xiofQ}]
at $Q_s^2(x_m)$ and $x_m=Q^2_s(x_m)/\Qtmax$, one obtains
\be
\label{longQs}
\ln\frac{Q^2_s(x_m)}{\mu^2}&=&\frac{1}{2}
\left[
\frac{2}{2+\blambda}\ln\frac{B}{T\mu^2}+\ln\frac{\Qtmax}{\mu^2} \right.
\nonumber \\
&&\hspace{-0.8in}+\left.
\sqrt{\frac{\blambda}{2+\blambda}\ln^2\frac{\Qtmax}{\mu^2}
-\frac{2 \blambda}{(2+\blambda)^2}\ln^2\frac{B}{T\mu^2}}
\right]\, .
\ee
In the large energy limit, the above solution simplifies to
\be
\label{xmdef}
\frac{Q^2_s(x_m)}{\mu^2}&\approx &
\left(\frac{B}{\mu^2 T}\right)^{\frac{1}{2+\blambda}}
\left(\frac{\Qtmax}{\mu^2}\right)^{\frac{1}{2}
+\frac{1}{2}\sqrt{\frac{\blambda}{2+\blambda}}}
.
\ee

To compute the quenching parameter we study numerically the 
integral in \Eq{qhatofx}
and find that, for the infinite conformal plasma, it can be well approximated by
\be
\label{qhatofxs}
\hat{q}_R&=&\frac{C_R}{N_c}\frac{Q^2_s(x_m)}{\min(L,L_c(x_m))}
\frac{2}{\pi}\alpha_s\left(Q^2_s(x_m)\right)
\nonumber \\
&\times&
 N_c \ydm \left[
\frac{\delta_L}{\sqrt{2\ydm\xi(Q^2_s(x_m))}} \right.\nonumber \\
&+& \left.
\frac{1}
{\xi(Q^2_s(x_m))-\frac{2}{\pi}N_c\alpha_s\left(Q^2_s(x_m)\right)\ydm}
 \right]\, .
\ee
This is a very good approximation for values of $\blambda>1$. For small 
$\blambda$ it approximates the exact integral within a factor 2 as long
as $\blambda \ln (\Qtmax/T^2)>1$.
Substituting \Eq{xmdef} in \Eq{qhatofxs} we find
\be
\label{qhatofqsat}
\hat{q}_R&=&\frac{C_R}{N_c} Q^2_s(x_m) T x_m
\left(\frac{\ydm}{\xidmx}
\right. \nonumber \\
&& \hspace{0.3in} + \left. \frac{1}{2}\sqrt{\frac{\blambda}{2}}
\frac{\ydm}{\sqrt{\ydm\xidmm}}
          \right) \, .
\ee
As expected \cite{Baier:2002tc}, the transport parameter is determined by the 
saturation scale.

To determine the dependence on the coupling, we substitute the
definition of $B$
and
set $\mu^2=T^2$
\be
\label{qhat_L}
\frac{\hat{q}_R}{T^3}&=&\frac{C_R}{N_c}
\left(\frac{Q^2_{max}}{T^2}\right)^{\sqrt{\frac{\blambda}{2+\blambda}}}
\left[\frac{\pi^{5/2}
      \blambda^{5/4}(2+\blambda)^{1/4}}
{36\sqrt{\ln\Qtmax/T^2}}
\right]^{\frac{2}{2+\blambda}} \nonumber \\
&&
\times\left(\sqrt{2+\blambda}-\sqrt{\blambda}\right)^{\frac{4+\blambda}{2+\blambda}}
\frac{1}{4}\left[
\sqrt{\blambda}+\frac{2}{\sqrt{\blambda}}
\right] \, .
\ee
Let us point out two interesting features in the above result: 1) $\hat{q}_R$
grows as a coupling dependent power of the energy and 2)
it depends non analytically on the reduced t'Hooft
coupling $\bar \lambda$. The non-analyticity is a consequence of
the evolution process.

\begin{figure}
\centerline{\includegraphics[width=8cm]{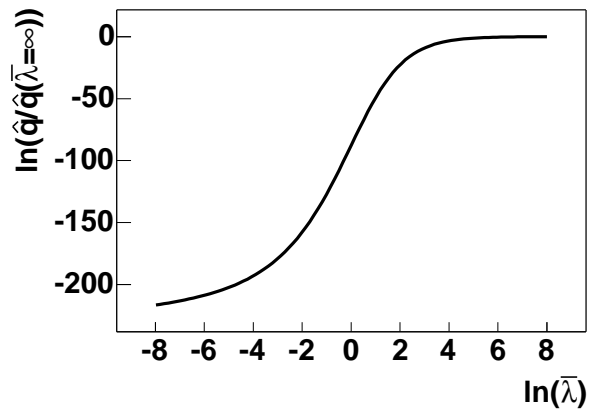}}
\caption{Normalized jet transport parameter as a function of
reduced t'Hooft coupling $\bar \lambda$ in a formal plasma for an energetic 
probe of $\log_{10}(\Qtmax/T^2)=90$.}
\label{fig:qofl}
\end{figure}

The derivation of \Eq{qhat_L} for a conformal plasma is strictly
valid for small values of $\blambda$ since both the evolution
equation \Eq{DLAeq} and the initial conditions \Eq{eq:g-g1} are
based on perturbation theory. However, in our computation we have
not make any further assumption about the smallness of $\blambda$.
Given the recent interest in the computation of transport
properties in strongly coupled $\mathcal{N}=4 $ SYM
\cite{Liu:2006ug,Gubser:2006nz,jorge:2007qw,
Herzog:2006gh,Casalderrey-Solana:2006rq}, it is still instructive
to study the strong coupling behavior of the jet transport
parameter. Plotted in Fig.~{\ref{fig:qofl}} is the jet transport
parameter as a function of  the reduced t'Hooft coupling $\bar
\lambda$, normalized by its large coupling limit, \be
\label{qstrong} \hat{q}_R(\blambda=\infty)=\frac{C_R}{N_c}\frac{T
Q^2_{max}}{4} =\frac{C_R}{N_c}  \frac{3T^2 E}{2} \, . \ee The
normalized jet transport parameter increases monotonically with
the coupling $\bar \lambda$ and reaches its asymptotic value in
the strong coupling limit. Note also that the above limit assumes that the 
energy of the probe is large such that $\ln (\Qtmax/T^2)>> \ln \blambda$. 

Several comments on this strong coupling limit are in order:
\begin{itemize}
\item In the strong coupling limit, the saturation scale
approaches its maximum limit $Q_s^2(x_m)=Q^2_{max}$ and, thus, the
eikonal approximation is questionable. Both the saturation scale
and the transport parameter become independence of the initial
condition as contained in $B$. \item \Eq{qstrong} has a power
dependence on the energy of the probe, and the power becomes
coupling independent in the strong coupling limit. \item
Contribution to \Eq{qstrong} comes completely from the saturated
part of the gluon distribution function. We have performed a
simplified treatment of this region by considering it constant.
This is well motivated by numerical solutions of the
Balitsky-Kovchegov equations at weak coupling
\cite{Albacete:2004gw,Albacete:2007yd}. However, at strong
coupling extra dependencies on the coupling (subleading at weak
coupling) may become important. \item We have not considered the
evolution of the wave function of the probe. This is motivated by
the weak coupling picture, in which such evolution is considered
separately as the radiative processes of the probe and are
described by radiative energy loss. At strong coupling this
separation of the probe and medium evolution becomes ambiguous and
may lead to an extra coupling dependence.
\end{itemize}

\section {Non-conformal Plasma}

From the analysis of a conformal plasma with a fixed coupling
we concluded that the saturation scale grows faster than any logarithmic
jet energy dependence. Since the typical momentum scale is
dictated by $Q^2_s$, effects of a running coupling constant
become important in the QCD plasma for large energy probes. This issue is
addressed by 
 solving numerically \Eq{qhatofx} with $\xi$ given by \Eq{xiofQ}. In this case
we find that \Eq{qhatofx} is well approximated (within 20 \%)
\footnote{
For determining this expression we assumed that the coherence length is always larger
or smaller than the path length. In the numerical computations presented this is not 
assumed and the $\min()$ is replaced by a smooth function.  
}
 by 
\be
\hat{q}_R&=&\frac{C_R}{N_c}\frac{Q^2_s}{\min(L,L_c)}
\frac{\ydm}{\ln{\frac{Q^2_s(x_m)}{\Lambda^2}}}
\nonumber \\
&&
\times\left[
\frac{\delta_L}{\sqrt{\pi\frac{b}{N_c} \ydm\lxinc}}\right. \nonumber \\
&&+ \left.
\frac{1}{\lxinc-\frac{\ln(1/x_m)}{\ln(Q^2_s(x_m)/\Lambda^2)}}
\right]\, ,
\label{Qsdefnc0}
\ee
where $\delta_L=1/2$ if $L>L_c$ and $\delta_L=1$ otherwise. As in the conformal
case, $\hat{q}_R$ is determined by the saturation scale,
which is given by
\be
\label{Qsdefnc}
Q^2_s&=&B(x,Q^2_s)\min(L,L_c) \nonumber \\
    &\times&e^{\sqrt{\frac{4 N_c}{\pi b}\lxinc \ydm}} \, ,
\ee
 where $B(x,Q^2_s)$ is given
in \Eq{Bdef} and $L_c=1/x_mT=Q^2_s(x_m)/\Qtmax T$.

As in the conformal case, the saturation scale and jet transport
parameter have a fast growth with the jet energy.
In spite of the fact that the above results are derived with
an approximation for asymptotically small $x$ (which
implies large saturation scales), we would like to make some
numerical evaluations of the jet transport parameter for jet
energies accessible at RHIC and LHC and address
 the experimental consequences of this growth.

We solve numerically the self-consistent equation \Eq{Qsdefnc} for
the saturation scale $Q_s^2(x_m)$. In order to avoid the infrared
singularity of $\alpha_s$ we regulate the coupling constant as \be
\alpha_s(Q^2)=\frac{1}{b}\frac{1}{\ln\frac{Q^2+T^2}{\Lambda^2}} \,
. \ee To be consistent, we also replace \be \lxinc \rightarrow
\lxincR \, . \nonumber \ee The coupling constant at a thermal
scale is determined by solving the coupled equations, \be
\alpha_s^T&=&\frac{1}{b}\frac{1}{\ln\frac{Q^2_s(T)+T^2}{\Lambda^2}}
\, ,
\\
Q^2_s(T)&=&18\pi\frac{(N_c \alpha_s^T)^2}{Q^2_s(T)}  T^4 \, ,
\ee
which are essentially \Eq{fullQsT} at $x=Q^2_s(T)/18T^2$ [we have set
the logarithms in \Eq{fullQsT} to be of order 1].
Finally, since \Eq{Qsdefnc} is only valid for asymptotically
large rapidities $y$, we also shift the rapidities to
\be
y \rightarrow y+y_0 \, ,
\ee
with $y_0 = 0.24$. This value has been chosen such that as $y$ decreases
we recover the value of the saturation scale $Q^2_s(T)$. In the
following numerical evaluation we will
choose $\mu^2=T^2$ and $\Lambda=200$ MeV.

\begin{figure}
\centerline{\includegraphics[width=8cm]{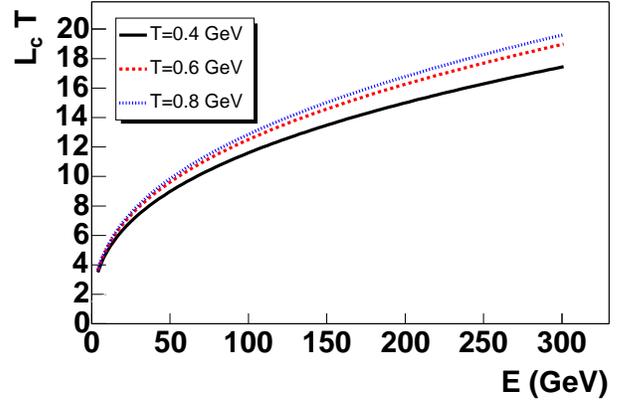}}
\caption{(Color online) Coherence length times temperature as a function of the
energy of the probe for different temperatures.}
\label{Lcplot}
\end{figure}

Since the medium is finite in heavy ion collisions, we start by
studying the coherence length. This is computed by solving \be
\label{Lc_E}
L^2_c&\equiv& \frac{1}{x_m ^2 T^2} \nonumber \\
&=& \frac{6 E}{B}e^{-\sqrt{\frac{4 N_c}{\pi b}\lxincm \ydm}} \, .
\ee This coherent length is used to calculate the saturation scale
for any large medium size $L>L_c$. For small medium size, $L<L_c$,
the actual length $L$ is used to calculate the saturation scale.
The coherence length is plotted in \Fig{Lcplot} and it shows a
strong energy dependence, as can be inferred from \Eq{Lc_E}. This
strong jet energy dependence $L_c \sim \sqrt{E} $ is approximately
independent of the evolution of the gluon distribution function,
and stems from the definition of the coherence length as
$L_c=1/xT$. As expected from the running coupling, it does not
scale with temperature. For a characteristic temperature of
$T=0.4$ GeV in relativistic heavy ion collisions, the coherence
length is significant: for a probe of $E=20$ GeV, $L_c\approx 2.5$
fm while at $E=100$ GeV, $L_c \approx 4.5 $ fm, which are
comparable with the nuclear size.

\begin{figure}
\begin{center}
\includegraphics[width=8cm]{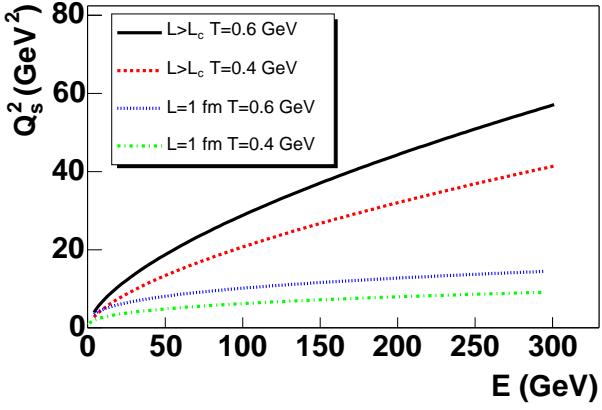}
\end{center}
\caption{(Color online) Saturation scale as a function of the jet energy.
}
\label{QshqLplot}
\end{figure}

\begin{figure}
\begin{center}
\includegraphics[width=8cm]{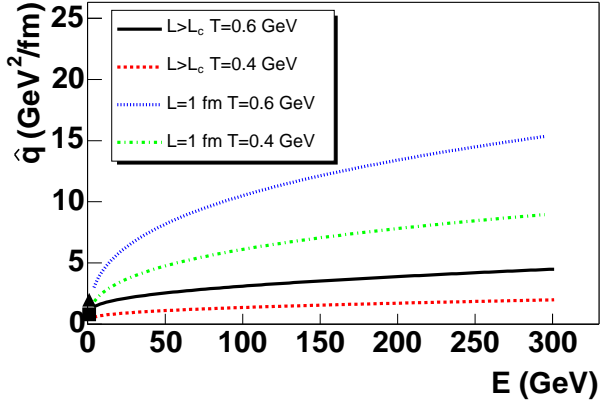}
\end{center}
\caption{(Color online) Jet quenching parameter $\hat{q}$
as a function of the jet energy.
The square (triangle) marks the the value of $\hat{q}$ for thermal particle 
at $T=0.4 $ GeV ($T=0.6 $ GeV).
Significant corrections to the energy dependence are expected
at low energy which should approach their thermal value at $E=3T$.}
\label{qhatLplot}
\end{figure}

When the coherence length becomes comparable to the medium size a
non-trivial length dependence of the saturation scale will arise,
since the definition of $Q^2_s$ is different for path lengths
longer or shorter than the coherence length. This is illustrated
in \Fig{QshqLplot} where the saturation scale is plotted as a
function of the energy of the probe.
 When the path length is longer than the coherence length,
$Q^2_s$ shows a stronger dependence on the energy. This is, in
fact, driven by the energy dependence of the coherence length, and
is mostly independent of the evolution of the medium gluon
distribution.
 When the path length is smaller than the coherence length
we obtain a significant reduction of the saturation scale and
a much weaker dependence on the energy, since the DLA evolution
leads to a growth that is weaker than a power but faster than a
logarithmic dependence.
We note that the gluon saturation scale obtained here for a gluonic
plasma is significantly larger than that in a nucleon,
where $Q^2_s\approx 1$ GeV$^2$ at $x\approx 10^{-4}$ \cite{Golec-Biernat:1998js}.
This is a consequence of the
fact that the QGP is a much denser system than a nucleon or
cold nucleus. The saturation scale in a heavy nucleus is
enhanced by a factor of $A^{1/3}$ and therefore might
be large enough to facilitate a perturbative calculation of
gluon distributions \cite{McLerran:1993ni}. However, it is still an order
of magnitude smaller than in a high temperature quark-gluon plasma.

In \Fig{qhatLplot} we show the value of the jet quenching
parameter $\hat{q}_R$ from the integration of \Eq{qhatofx}. For long path lengths $\hat{q}$
becomes path length independent. The leading energy dependencies
of both $Q^2_s$ and $L_c$ cancel and the observed energy
dependence is a consequence of the evolution of the medium gluon
distribution. At shorter path lengths, we obtain an enhancement of
$\hat{q}_R$ as a consequence of the evolution.

\begin{figure}
\begin{center}
\includegraphics[width=8cm]{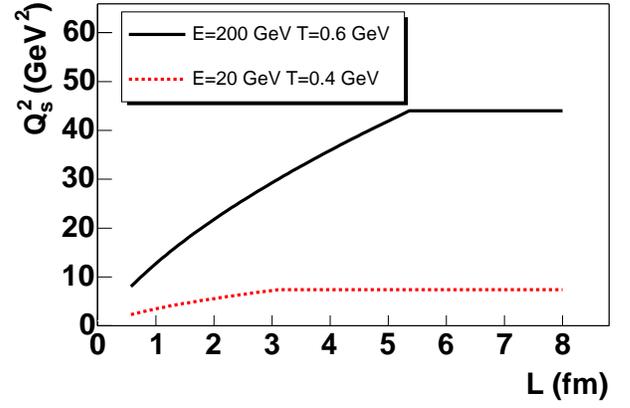}
\end{center}
\caption{(Color online) Saturation scale as a function of the path length for
different
 probe energies.
}
\label{QshqofLplot}
\end{figure}

\begin{figure}
\begin{center}
            \includegraphics[width=8cm]{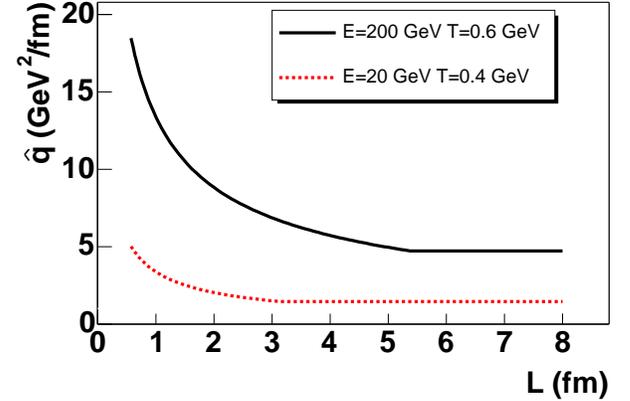}
\end{center}
\caption{(Color online) Jet quenching parameter $\hat{q}$ as a
function of the path length for different
 probe energies. 
 }
\label{hatqofLplot}
\end{figure}

The length dependence of both the saturation scale and the
transport parameter is shown in Figs. \ref{QshqofLplot} and
\ref{hatqofLplot} for different energies of the probe. As shown in
\Fig{QshqofLplot}, the saturation scale grows as a function of the
path length. However this growth is smaller than linear. Thus, the
transport parameter, as shown in \Fig{hatqofLplot}, diverges at
small path length. Note, however, that at very short path lengths
($L\ll \lambda_{f}$) the mean momentum broadening should vanish,
since the probe has no medium to scatter with. Thus, we expect
correction to the small $L$ dependence of both $Q^2_s$ and
$\hat{q}_R$. Finally, when the path length is larger than the
coherence length both quantities become length independent. Note
that we have assumed a simplified transition from the region
$L<L_c$ to $L>L_c$. This is the reason for the abrupt change in
the length dependence at $L=L_c$ in  Figs.~\ref{QshqofLplot} and
\ref{hatqofLplot}.

Let us remark that for path lengths smaller than the coherence length,
the interaction of the probe with the whole length of the medium is coherent. 
Thus, if the length scales of space and time variation are smaller or 
comparable to the path length, the analysis of the saturation scales and
jet transport parameter should be revisited. This will
complicate the phenomenological extraction of the transport 
parameter in an expanding medium with strong spatial variation
as in semi-peripheral heavy-ion collisions.

\begin{figure}
\begin{center}
\includegraphics[width=8cm]{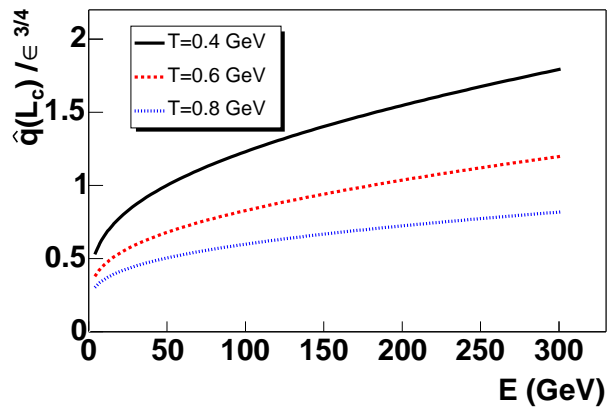}
\end{center}
\caption{(Color online) Jet quenching parameter scaled by $\epsilon^{3/4}$, with
$\epsilon$ the energy density.}
\label{Kplot}
\end{figure}

Because of the running of the coupling constant or the intrinsic
scale ($\Lambda$) in QCD as a non-conformal gauge theory, the
transport parameter $\hat{q}_R $ has a non-trivial temperature
dependence. To illustrate this, we plot in \Fig{Kplot} the value
of $\hat{q}_R$ scaled by the energy density, \be
\epsilon=\frac{8\pi^2}{15}T^4 \, , \ee to the power of $3/4$ for
the long path lengths ($L>L_c$). The dependence on the temperature
is quite strong in the temperature range showed. This is not
surprising since these temperatures are of the order of $\Lambda$
and the coupling constant is very sensitive to the scale. The
dependence, however, becomes weaker at higher temperatures. The
jet energy dependence of the transport parameter is also stronger
at lower temperatures.  This, of course, is only a
lower limit, since $\hat{q}_R$ is larger for shorter 
path lengths~\footnote{The initial value of $\hat{q}_R$
before evolution is about half that of \cite{Baier:2006fr}. The
main reason is that we use $\alpha_s=0.3$ for $T=0.4$ GeV while in
\cite{Baier:2006fr}, $\alpha_s=0.5$.}.

\section{Bound on $\hat q$ and shear viscosity to entropy density ratio}

Following Ref.~\cite{Majumder:2007zh}, one can relate the jet
quenching parameter $\hat q_R$ to the transport mean-free-path
of the hard probe,
\begin{equation}
\label{lmfp}
\lambda_{f}^{-1}\approx\frac{4\rho}{\langle \hat s\rangle}
\int dq_T^2 q_T^2\frac{d\sigma_R}{dq_T^2}
=\frac{4\hat q_R(E)}{\langle \hat s\rangle},
\end{equation}
where we have used the definition of jet transport parameter in
\Eq{eq:qhat0} and $\langle \hat s\rangle=Q_{max}^2\approx 6ET$
is the average center-of-mass energy squared
of the jet-gluon scattering. The requirement that the mean-free-path of
the hard probe must be larger than the
de Broglie wave length $1/E$ will set
an upper bound for the energy loss parameter,
\begin{equation}
\hat q_R (E)\leq \frac{\langle \hat s\rangle E}{4}=\frac{3 E^2 T}{2 C } \, ,
\label{bound1}
\end{equation}
where $C$ is a constant on the order of ${\cal O}(1)$.

\begin{figure}
\begin{center}
\includegraphics[width=8cm]{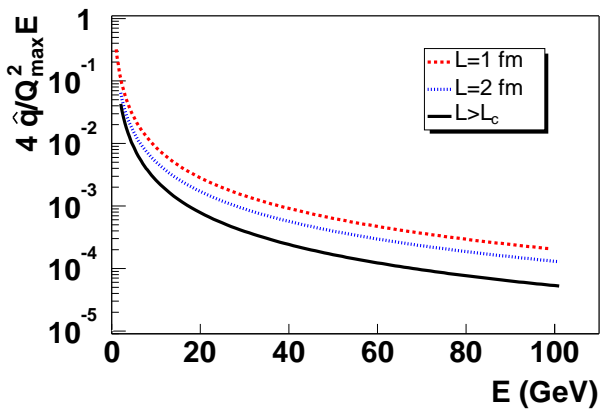}
\end{center}
\caption{(Color online) Ratio $4\hat{q}/Q_{max}^2E=2\hat{q}/3E^2T$ as a function of jet energy $E$.}
\label{fig:bound1}
\end{figure}

We have checked that our numerical
solutions of $\hat q_R$  indeed satisfy this condition as shown in
Fig.~{\ref{fig:bound1}}. For a conformal plasma, $\hat q_R$
in \Eq{qhat_L} increases monotonically
with the reduced t'Hooft coupling $\bar \lambda$(see Fig.~\ref{fig:qofl}).
In the
limit  $\bar \lambda\rightarrow \infty$, $Q_s^2(x_m)=Q^2_{max}$
and the jet transport parameter $\hat q_R=C_R T Q^2_{max}/4N_c$
for a gluon jet satisfies the bound for $E\geq T$.
Since the strong coupling limit is an asymptotic behavior, the
above bound on the transport parameter is also satisfied in the
weak coupling limit of a conformal plasma.

For thermal partons with $\langle E\rangle \sim 3T$,
the upper bound in \Eq{bound1} becomes
\begin{equation}
\frac{T^3}{\hat q_R(T)}\geq \frac{2}{27C}\, .
\label{eq:bdqhat2}
\end{equation}
According to Ref.~\cite{Majumder:2007zh}, one can also relate the
shear viscosity to the transport parameter for a thermal parton,
\begin{equation}
\eta\approx \frac{1}{3}sT\lambda_{f}\approx s\frac{3T^3}{2\hat q_R(T)},
\,\,\,\, {\rm or}\,\,\,\,
\frac{\eta}{s}\approx \frac{3}{2}\frac{T^3}{\hat q_R(T)}\, .
\label{eq:etaqhat}
\end{equation}
Therefore, the upper bound on the transport parameter $\hat q_R(T)$ also
provides a lower bound on the shear viscosity to entropy density ratio
\begin{equation}
\frac{\eta}{s}\geq \frac{1}{9C}\, .
\end{equation}
This is very similar to the bound found by Danielewicz and
Gyulassy \cite{Danielewicz:1984ww} from transport theory and the bound
$1/4\pi$ found in the strong coupling limit of
$\mathcal{N}=4 $ SYM \cite{Policastro:2001yc}.

 The upper bound on the transport parameter $\hat q_R$ in
\Eq{bound1} and its connection with the shear viscosity in
Eq.~(\ref{eq:etaqhat}) are quite general since they do not rely on
any assumption about the nature (perturbative or nonperturbative)
of parton interaction inside the medium. 
They do rely, however, on a transport description of the  
 plasma in terms of quasi-particles. Therefore,
it is not surprising that the connection between the transport
parameter and shear viscosity does not hold in the strong coupling
limit of $\mathcal{N}=4 $ SYM theory, since thermal modes in such
a strongly coupled theory cannot be described as quasi-particles
\cite{Teaney:2006nc}.

We should emphasize that the jet transport parameter is an
intrinsic property of the medium which could be dominated by
non-perturbative physics. However, in the case of large saturation
scale $Q_s^2$ and transverse momentum transfer,
the evolution of the gluon distribution function and the jet
transport parameter should still be described by perturbative QCD
and so should the interaction between jet and the medium and the
radiative energy loss. Therefore, as far as the transport
description of the dense medium is valid, one can use
the energy dependence of the jet (parton) transport parameter
$\hat q_R(E)$ as determined by jet quenching phenomenology to
estimate the shear viscosity to entropy density ratio via
extrapolation.

With a recent phenomenological study of both single and dihadron
spectra suppression using the next-to-leading order (NLO) pQCD
parton model calculation \cite{Zhang:2007ja}, the average gluon jet
energy loss per unit length in a 1-d expanding medium is estimated
to be
\begin{equation}
(\frac{dE}{dL})_{1d}\approx 1.9 - 3.4 \,\,\,{\rm GeV/fm},
\end{equation}
for $E=10-15$ GeV, which also includes an empirical variation with
jet energy. Using the relationship between parton energy loss
and the transport parameter in Eq.~(\ref{bdmps}), one obtains
an estimate of the average gluon jet transport parameter
\begin{equation}
\hat q_0(E)=\frac{2}{\tau_0 \alpha_s N_c}(\frac{dE}{dL})_{1d}
\approx 1.0 - 1.9 \,\, {\rm GeV}^2/{\rm fm}
\end{equation}
at an initial time $\tau_0=1$ fm/$c$ \footnote{This phenomenological
analysis was based on the assumption that $\hat{q}$ is independent
of the path length. The nontrivial length dependence of
$\hat{q}$ as obtained in this paper will affect the extracted average
value of the jet transport parameter.}. 
Here, we used $\alpha_s\approx 0.24$
at $Q_s^2\approx 5$ GeV$^2$ in a pure gluonic plasma.
This roughly agrees with the numerical calculation in Fig.~\ref{hatqofLplot}.
As shown in Fig.~\ref{qhatLplot}, the energy dependence of
$\hat q_R$ is very weak for $E<20$ GeV and long propagation length.
Even for short propagation length $L\sim L_c\sim 2$ fm, the energy
dependence is limited to about 25\% variation. We therefore can use
the above estimate for thermal parton transport parameter.
Using the same initial temperature $T_0=(337\pm 10)$ MeV as in
Ref.~\cite{Majumder:2007zh}, one obtains
\begin{equation}
\frac{\eta_0}{s_0}\approx \frac{3}{2}\frac{T_0^3}{\hat q_0(T)}
\approx 0.15 - 0.24.
\end{equation}
For more consistent analysis, one should consider explicit energy
dependence of the parton energy loss beyond that in $\hat q_R$.

Finally, the requirement that the coherence length is larger than
the mean-free-path of thermal particle sets a limit on the coupling
constant $\alpha^T_s$. Indeed, from  \Eq{lmfp} and \Eq{Qsdefnc0}
$\hat q_R\approx x_mQ_s^2(x_m)T$,
we find
\begin{equation}
x_m=Q_s(x_m)/\langle \hat s\rangle\leq 1/4\, .
\end{equation}
This bound is only satisfied at weak coupling
\begin{equation}
2N_c\alpha^T_s\sqrt{2\pi\ln\frac{1}{N_cg^2}}\leq 1 \, .
\end{equation}
For larger values of the coupling $\alpha^T_s$, coherence effects in the
multiple scattering of thermal particles become important.

\section{Summary and Discussions}

In this paper we have studied the energy dependence of the jet
transport (or quenching) parameter $\hat{q}_R$. By relating
$\hat{q}_R$ to the unintegrated gluon distribution function of the
plasma we have shown that the energy dependence of $\hat q_R$
arises from the evolution of the gluon distribution function.
Thermal quarks and gluons in the plasma play the same role as the
valence quarks of the nucleus and high energy jets probe their
wave functions at small $x$. Similar as in a cold nuclear matter,
the evolution leads to a growth in the gluon number which is
eventually tamed by saturation effects. Therefore, the jet
transport parameter, also defined as the momentum broadening per
unit length, is determined by the saturation scale $Q^2_s$.

Using thermal field theory with HTL resummation, we have derived
the gluon distribution function
for scales $\mu^2<T^2$ as probed by the interactions among
thermal partons. For such interaction among thermal partons,
the coherence length is smaller than the mean-free-path
and, thus, the saturation scale grows fast for small $x$, $Q^2_s\sim1/x$.
Remarkably, evaluating the saturation scale at the typical $x$ of
the scattering among thermal partons, $x\approx Q^2_s/4T^2$, leads
to $Q^2_s\sim \mu^2_D$. Therefore, the
typical momentum transfer is of the order of $\mu_D$ as expected.
What is more interesting is that, for large angle scattering of
thermal particles with $x\sim 1$, the saturation
scales is of the order of the magnetic mass $Q^2_s\sim \mu^2_{mag}$.

The hard thermal loop result for gluon distribution function
serves as the initial condition of the evolution of the
gluon distribution as probed by an energetic jet. Since this
process involves scales much larger
than the medium scale, we have neglected thermal modification
of the evolution kernel. We then used the double logarithmic
approximation to describe the evolution in a conformal theory
and in (pure glue) QCD.

For a conformal plasma, both the saturation scale and the jet
transport parameter $\hat{q}_R$ grow with energy as a
coupling-dependent power. The evolution leads to a $\hat{q}_R$
which is non-analytic in the t'Hooft coupling $\lambda=g^2 \,N_c$.
In the large coupling limit, $\hat{q}_R$ becomes independent of
the coupling and grows linearly with E. This is very different
from results obtained in $\mathcal{N}=4$ SYM
\cite{Liu:2006ug,Herzog:2006gh,Casalderrey-Solana:2006rq,
Gubser:2006nz,jorge:2007qw}. As remarkable as this may be, the
analysis we have performed is perturbative in nature and the
extrapolation to infinite coupling might not be justified.

In the case of (pure glue) QCD, the evolution leads to a jet
energy dependence of the transport parameter that is stronger than
any power of logarithmic dependence. The saturation effect also
gives rise to a non-trivial length dependence of the jet transport
parameter.  The running of coupling constant also results in a
significant temperature dependence which becomes weaker at higher
temperatures. We have numerically evaluated the saturation scale
and jet transport parameter in a temperature range $T=0.4-0.6$ GeV
that is relevant for relativistic heavy ion collisions at RHIC and
LHC. The growth of $\hat{q}_R$ with jet energy
is modest for large medium size $L\gg L_c$. However, the
energy dependence is significant for $L \stackrel{<}{\sim} L_c$. 
The obtained transport parameter is also
larger than that computed via perturbation theory without
evolution \cite{Baier:2006fr}. For $T=0.4$ GeV and $E=10-20$ GeV,
our computed value $\hat q_R\approx 1.5-2$ GeV$^2$/fm for a gluon jet
is in agreement with the results from recent phenomenological
studies of experimental data on jet quenching at RHIC
\cite{Turbide:2005fk,majumder,Zhang:2007ja,Majumder:2007ae}. 
It is, however, smaller than the results
from phenomenological studies \cite{Eskola:2004cr,Dainese:2004te} 
based on an implementation of energy loss by
Salgado and Wiedemann (SW) \cite{Salgado:2003gb}.
A recent study within this model with explicit space-time dependent
profiles of energy density from 2 and 3-D hydro calculations gives
a transverse averaged $\hat q_R\approx 4-5$  GeV$^2$/fm \cite{Renk:2006pk}
at initial time $\tau_0=1$ fm/$c$. 
Inclusion of dihadron suppression
in the phenomenological study has been shown \cite{Zhang:2007ja}
to greatly improve the sensitivity of jet quenching to
variation of $\hat q_R$. It is clear that inclusion of
the energy and length dependence of the jet transport parameter 
will also influence the phenomenological study of the jet quenching
measurements.

Given the relation between shear viscosity $\eta$ and transport
parameter $\hat q_R(T)$ for a thermal parton as derived recently
in Ref.~\cite{Majumder:2007zh}, the energy dependence of jet
transport parameter determined from theoretical and
phenomenological studies can also be used to estimate the shear
viscosity in the dense matter produced in heavy-ion collisions.
This will unify high $p_T$ and low $p_T$ aspects of heavy-ion
collisions. The latter can characterize the collective behavior of
the produced dense matter which is well described by relativistic
hydrodynamics with a negligible shear viscosity
\cite{Teaney:2003kp}. We have also derived an upper bound on the
transport parameter of high energy jets. This upper bound can
lead to a lower bound on the shear viscosity to entropy density
ratio which is consistent with other transport studies.

The saturation scale in a glue plasma we obtained in this study is
much larger than that in a cold nucleus or a nucleon at extremely
small $x$. This results from the high gluon density in a plasma
with coherence length comparable to the medium size. This is
quite different from the analysis of Hadron-Electron Ring 
Accelerator (HERA) data which leads to a power 
dependence of $Q^2_s$ on $x$ \cite{Golec-Biernat:1998js}
\be
\label{HERAQs}
Q^2_s=1 \, {\rm GeV} \, \left(
                       \frac{3 \times 10^{-4}}{x}
                       \right)^{0.288} \, ,
\ee where the coherence length is much larger than the nucleon
size. When the coherence length is comparable to the medium
length, the saturation scale is not linear with the path length.
Thus, instead of using the phenomenological expression
\Eq{HERAQs}, we used the DLA approximation to estimate $Q^2_s$.
Numerical solutions to the Balitsky-Kovchegov (BK) equation  with
running coupling constant \cite{Albacete:2004gw,Albacete:2007yd}
show that the unintegrated gluon distribution is consistent with
the DLA asymptotes and that the saturation scale behaves as
$Q^2_s\sim \exp\{\Delta\sqrt{y}\}$, similar to what is expected
from the DLA asymptotes. Note, however, that these conclusions are
for large values of the rapidity, $y\sim 10$, while in our case
the typical $x$ probed by high energy jets in the kinematic range
relevant to heavy-ion collisions at RHIC and LHC is not very
small, $x \sim 0.1$.  The large values of $x$ obtained imply that
significant corrections to the obtained behavior may occur, which
could be addressed by a numerical analysis of the BK equation.

{\it Acknowledgments:} This work was supported in part by the
U.~S.~Department of Energy under grant
contract DE-AC02-05CH11231. We thank J. L. Albacete, N. Armesto,
P. Jacobs, C. Salgado, D. Teaney, T. Renk and U. Wiedemann
 for fruitful discussions.

\end{document}